\documentclass[preprint,showpacs]{revtex4}
\usepackage{graphicx} 
\begin{document}
\title{Crossover of interface growth dynamics during corrosion and passivation}
\author{ F. D. A. Aar\~ao Reis${}^{1,}$\footnote{Email address:
reis@if.uff.br} and
Janusz Stafiej${}^{2,}$\footnote{Email address:
janusz\_stafiej@yahoo.com}
}
\affiliation{
${}^{1}$ Instituto de F\'\i sica, Universidade Federal Fluminense,\\
Avenida Litor\^anea s/n, 24210-340 Niter\'oi RJ, Brazil\\
${}^{2}$ Institute of Physical Chemistry, Polish Academy of Sciences,\\
Kasprzaka 44/52, 01-224, Warsaw, Poland
}
\date{\today}

\begin{abstract}

We study a model of corrosion and passivation of a metalic surface in contact
with a solution using scaling arguments and simulation. The passive layer is
porous so that the metal surface is in contact with the solution. The volume
excess of the products may suppress the access of the solution to the metal
surface, but it is then restored by a diffusion mechanism. A metalic
site in contact with the solution or with the porous layer can be passivated
with rate $p$ and volume excess diffuses with rate $D$. At small times, the
corrosion front linearly grows in time, but the growth velocity shows a
$t^{-1/2}$ decrease after a crossover time of order $t_c\sim D/p^2$, where the
average front height is of order $h_c\sim D/p$. A universal scaling relation
between $h/h_c$ and $t/t_c$ is proposed and confirmed by
simulation for $0.00005\leq p\leq 0.5$ in square lattices. The roughness of the
corrosion front shows a crossover from Kardar-Parisi-Zhang scaling to Laplacian
growth (diffusion-limited erosion - DLE) at $t_c$. The amplitudes of roughness
scaling are obtained by the same kind of arguments as previously applied to the
other
competitive growth models. The simulation results confirm their validity. Since
the proposed model captures the essential ingredients of different corrosion
processes, we also expect these universal features to appear in real systems.

\end{abstract}

\pacs{}
\maketitle

\section{Introduction}
\label{intro}

The corrosion of metalic surfaces is a subject of increasing interest due to
widespread use of corrodable materials with all the economical and environmental
implications \cite{Smialowska, Frankel}.
When the metal is in contact with an aggressive solution, the formation of a
passive layer containing the metal oxide is observed after the corrosion of the
topmost layer of the surface.
As this layer grows, it more efficiently protects the surface from the attack
of the solution, leading to the slowdown of the corrosion rate. For instance,
according to the Cabrera and Mott\cite{cabrera} law, the corrosion front is
expected to move as $t^{1/2}$ (growth rate decreasing as $t^{-1/2}$), which is
in agreement with experimental findings.

The changes in the corrosion rate are certainly accompanied by changes in the
roughness of the front separating the metal and the passive
layer. In the beginning of the process, all the surface is uniformly exposed to
the corrosive solution, but after the formation of the passive layer there
appear points of easier or more difficult access to the solution. It is
particularly important to investigate the scaling properties of the roughness
because they are intimately connected to the main dynamical processes during
the interface evolution \cite{barabasi,krugreview}. From the
theoretical point of view, additional motivation comes from the renewed
interest in systems whose competitive dynamics leads to crossover in roughness
scaling \cite{albano1,shapir,tales,chamereis,lidia,kolakowska1,rdcor}.

Here we will analyze the crossover in the front evolution and in
the roughness scaling of a recently introduced model for
the corrosion of a metalic surface by a solution
\cite{saunier1,taleb,saunier2}. The basic mechanisms of the model are the
corrosion of the mettalic surface exposed to the solution and the formation of
the passive layer with the oxides. The layer is assumed to be porous, but the
larger molar volumes of the products may block the pores and, consequently,
diffusive relaxation of the volume excess is necessary for new corrosion
events. This model was not devised to represent a particular real process
\cite{saunier1}, but to contain essential ingredients of a large variety of
corrosion processes, allowing the investigation of universal features.

Previous numerical work on this model has shown significant differences in the
corrosion fronts for small and large values of the local corrosion rate of the
metal in contact with the solution \cite{saunier1}. However,
besides these qualitative features, only universal laws for the global growth
rates in limiting cases were already derived \cite{saunier2}.
In the present paper, we advance the quantitative study
of the model, focusing on the properties of the corrosion front 
which separates the metal and the passive layer.
We show the emergence of a crossover in the global growth rate accompanied by a
crossover in surface roughness scaling, from a short time KPZ behavior to long
time Laplacian growth (or diffusion-limited erosion - DLE) \cite{deutch,krug}.
A scaling picture of the process is derived, along the same lines of those
developed for related systems \cite{rdcor}, and the results are confirmed by
numerical simulation.

The rest of this paper is organized as follows. In Sec. II we present the model
for corrosion and passivation. In Sec. III, we present a scaling theory for the
front growth and numerical results that confirm it. In Sec. IV, we discuss the
crossover KPZ-DLE of the surface roughness. In Sec. V we summarize our results
and present our conclusions.

\section{The model of corrosion and passivation}

The model represents a system with the structure shown in Fig. 1a.
While the metal is corroded, a porous passive layer grows, with some 
products filling their pores. These products block the access of the solution
to the surface, but they may diffuse and leave the surface of the metal exposed
again.

For simplicity, the model is defined in a lattice, where the sites may have
three different states: M (metal), S (solution) or P (passive). The lattice
sites represent the dominant chemical species inside one a mesoscopic region.
Due to computational limitations of memory and time, a square lattice will be
considered in our numerical work. In order to represent the effect of the
porosity of the passive layer, we assume that a site M can be corroded
(passivated) even when it is in contact only with P sites, thus avoiding the
complication of representing the porous structure of the passive layer.
The excess of products will be represented by particles W, which can only lay at
P sites and may diffuse through the P matrix. At most one particle W is
allowed at each P site. A region of the lattice is illustrated in Fig. 1b.

In the beginning of the process, all sites above a certain row of the lattice
(height $h=0$) are labeled S, and all sites below or at that row are labeled M.
An M site is said to be in contact with the solution if it has at least one
neighbor S or one neighbor P without a W particle over it. At each time unit,
each M satisfying this condition may be changed into
P with probability $p$ (reaction $M\to P$). It corresponds to the reduction of
a chemical species in the solution by the metal and the formation of an
insoluble film. A reaction $M\to P$ is illustrated for the central site in Fig.
2a. On the other hand, Fig. 2b shows a case where this reaction is not possible
for the central M site due to the blocking effect of the W particle.

We assume that the molar volume of the oxide products is larger than that
of the metal (Pilling and Bedworth factor larger than one \cite{pilling}),
and we represent this excess by creation of W particles after the reaction
$M\to P$. Two W particles are produced immediately after each reaction: the
first one lays at the newly created P site and the second one lays at a
neighboring P or S site. The creation of the new W particles is also illustrated
in Fig. 2a. Some differences between models with creation of one or two W
particles per reaction were discussed in Refs.
\protect\cite{saunier1,taleb,saunier2}.

The particles W may also diffuse through the P layer. This represents
some type of stress relaxation which leads to the redistribution of the excess
of corrosion products \cite{saunier2}. At each time
unit, each W attempts to execute $D$ random displacements to neighboring sites.
The movement is possible only if the target site is P or S and if it does not
contain a W particle, otherwise the attempt is rejected. An acceptable movement
is illustrated in Fig. 2c, but an attempt to move the same W particle to any
other direction would be rejected in that case.

When a W particle attempts to move to an S site, or
when the W particle is created over an S site (top of the passive layer), this S
site is converted into P. This condition is reasonable to represent the
expansion of the passive layer towards the solution side. However, it does not
have a significant influence on the corrosion front, which separates the metal
and the passive layer, and which is the main interest of the present study. 

\section{Scaling of the height of the corrosion front}
\label{heightscaling}

Here we focus on the case of small corrosion rates $p$ and of not very small
values of $D$.

First we consider the short time behavior of the system.
In one time unit, the number of corroded sites is small compared to the total
number of sites in the front, but all W particles attempt to move during the
same time interval. The diffusion of W particles removes the restrictions for
the corrosion and the situation in Fig. 2b is rare: the W particle at the
neighboring P site will move 
away long before a successful corrosion attempt occurs at the central site.
Consequently, we expect that all M sites at the interface are
available for corrosion with equal probabilities. The distance of the
corrosion front from the initial level, hereafter called the height of the
corrosion front, increases as
\begin{equation}
h_{ini}\sim pt 
\label{hini}
\end{equation}
($h_{ini}$ is being measured in lattice units).
In this regime, the model is equivalent to an inverse of the Eden
model \cite{eden,barabasi}, in which a new particle is added to a randomly
chosen neighbor of the aggregate surface at each time step.

Now consider the system at long times. Since W particles move diffusively, their
upwards displacement away from the corrosion front increases as
${\left( Dt\right)}^{1/2}$, thus their velocity in the upwards direction
evolves as $\frac{d}{dt}\left[{\left( Dt\right)}^{1/2} \right] \sim
{\left( D/t\right)}^{1/2}$. This decreasing velocity leads to the accumulation
of W particles near the surface, which block new corrosion
events. This is schematically illustrated in Fig.3, where a sea of W particles
is shown at the bottom of the passive layer. 

The crossover from the initial regime to the long time one takes place at a
characteristic time $t_c$ in which the height of the sea
of W particles covers a finite fraction of the passive layer. The
height of that sea increases as
\begin{equation}
h_{W}\sim {\left( Dt\right)}^{1/2}
\label{hw}
\end{equation}
because it is determined by the diffusion of W particles produced at
short times. On the other hand, the height of the passive layer increases as
Eq. (\ref{hini}) before the crossover. Matching these heights we obtain the
crossover time
\begin{equation}
t_c \sim D/p^2 .
\label{tc}
\end{equation}
The height of the corrosion front at the crossover, measured in lattice units,
is
\begin{equation}
h_c \sim D/p .
\label{hc}
\end{equation}

After the crossover, a corrosion event is possible only when an M site has an
empty neighbor P, which requires the migration of voids (holes) through
the sea of W particles and their arrival at the corrosion front. When the M site
is corroded (reaction $M\to P$), the hole of the W sea disappears because it is
occupied by one of the new W particles. Since these holes move diffusively from
the top of the W sea to that front (similarly to the W particles), we conclude
that that the corrosion front will also move diffusively, with displacement
increasing as $t^{1/2}$ and velocity decreasing as $t^{-1/2}$.

The time evolution of the height $h$ of the corrosion front is expected to
follow a scaling relation that involves $h_c$ and $t_c$:
\begin{equation}
h/h_c = f\left( t/t_c\right) ,
\label{scalingh}
\end{equation}
where $f$ is a scaling function. Indeed, with $f(x)\sim x$ for small times, we
regain Eq. (\ref{hini}), and with $f(x)\sim x^{1/2}$ for long times we obtain
$h\sim {\left( Dt\right)}^{1/2}$, in agreement with the above discussion. 

In order to test the scaling relation (\ref{scalingh}), we use simulation data
for several values of $p$,
ranging from $p=5\times {10}^{-5}$ to $p=0.5$, and $D=1$. These data were
obtained in lattices with lateral length $L=10000$ and periodic boundary
conditions. In Fig. 4 we show a scaling plot of $hp$ versus $tp^2$, which are
respectively proportional to $h/h_c$ and $t/t_c$.
The excellent data collapse of the results for consecutive values of $p$ shows
that there is a universal relation between those quantities. Lines with slopes
$1$ and $1/2$ are also included in the plot in order to highlight the crossover
from the $h\sim t$ growth to the $h\sim t^{1/2}$ growth.

In previous numerical works with small and large $p$ \cite{taleb,saunier2}, this
universal behavior was not observed because, for a single value of $p$, it is
difficult to span a time interval long enough to show both scaling regimes. 
Numerical studies in three dimensions would certainly be more difficult.
However, the scaling picture will be the same in three dimensions because it is
derived from diffusion properties and the same stochastic rules for the model. 

\section{Scaling of surface roughness}
\label{roughness}

For small times, the equivalence to the Eden model (Sec. III) indicates that
the corrosion model is in the KPZ class of growth \cite{barabasi,
kpz}, i. e. in the continuous limit it is
represented by the KPZ equation 
\begin{equation}
{{\partial h}\over{\partial t}} = \nu{\nabla}^2 h + {\lambda\over 2}
{\left( \nabla h\right) }^2 + \eta (\vec{x},t) .
\label{kpz}
\end{equation}
Here, $h$ is the height at the position $\vec{x}$ in a
$d$-dimensional substrate at time $t$, $\nu$ and
$\lambda$ are constants and $\eta$ is a Gaussian
noise~\cite{barabasi,kpz}.
In the growth regime, the lateral correlation length is much smaller than
the lattice length and the roughness increases as $W\sim At^\beta$, where
$\beta$ is called the growth exponent ($\beta=1/3$ in $d=1$ \cite{kpz},
$\beta\approx 0.23$ in $d=2$ \cite{kpz2d}) and $A$
is an amplitude related to the coefficients of the KPZ equation.

In the original Eden model, each site of the front can grow with
probability $1$ at each time unit \cite{eden,barabasi}. However, the
corrosion of an M site in our model takes place with probability $p$ per time
unit. Thus, for small times we expect that
\begin{equation}
W_{KPZ}\approx A_{Eden} {\left( pt\right)}^\beta ,
\label{wkpz}
\end{equation}
where $A_{Eden}$ is a constant related to Eden growth properties. In Fig. 5 we
show a log-log plot of the small time roughness as a function of $pt$, for
small $p$, confirming the validity of Eq. (\ref{wkpz}). The average slope of
this plot is significantly below the
asymptotic KPZ value $\beta=1/3$, but these short time deviations are expected
in the Eden model \cite{wolf,kertesz} as well as in related KPZ models
\cite{balfdaar}.

At long times, a corrosion event takes place when a void of
the W sea arrives at the metal surface, after performing a random walk which
begins at a distant point. Consequently, surface
peaks will be more easily subject to corrosion than the valleys of the surface,
which favors the production of surfaces with low roughness, as observed in
previous numerical works for high values of $p$ \cite{saunier1,taleb}.
These features indicate that the long time regime is equivalent to that of the
DLE model \cite{deutch}. In
DLE, eroding particles are randomly left at points very distant from the
solid surface, they are allowed to diffuse and, when they reach the surface,
they are annihilated together with the particle of the solid. In our model, the
eroding particles of DLE are represented by the holes of the W sea.

In the continuous description of DLE, the stable interface is driven by the
gradient of a Laplacian field \cite{krug}. A logarithmic scaling of the squared
roughness is predicted in $1+1$ dimensions:
\begin{equation}
{W^2_{Laplacian}} \approx \frac{\Delta}{2\pi v}\ln{\left( 4\pi vt/a\right)} ,
\label{wLaplacian}
\end{equation} 
where $v$ is the growth velocity, $t$ is the time, $a$ is the lattice
parameter and $\Delta = v{a_r}^d$, where $a_r$ is a lengthscale associated with
the noise \cite{krug}.

Our model is equivalent to a DLE with a probability $p$ of the incident particle
to remove a
particle of the solid. Moreover, changes in the time scale are also expected
because the growth rate of our model decreases in time, while it is constant in
the original DLE \cite{deutch,krug}. Thus, for a connection of the DLE
roughness in Eq. (\ref{wLaplacian}) and that of our
model, we use $h=vt$ as the height of the corrosion front and $a=1$. This gives
\begin{equation}
{W^2_{DLE}} \approx B\ln{\left( 4\pi h\right)} ,
\label{wdle}
\end{equation} 
where $h$ is measured in lattice units and $B$ is a time-independent constant
associated with the noise ($p$ and $D$).

The time scaling of the surface roughness, implicitly given in its relation
with the height $h$ in Eq. (\ref{wdle}), is tested in Fig. 6, where we plot
$y=W^2/\ln{\left( 4\pi h\right)}$ as a function of $tp^2$. Large
values of $tp^2$ were chosen in order to ensure that only data in the long time
regime were considered ($tp^2>1$ - see Fig. 4), but simulations for extremely
long times are feasible only for large $p$. We notice that, for
a given value of $p$, $y$ oscillates around some constant value, which provides
the estimate of the constant $B$ in Eq. \ref{wdle}. 

The crossover from KPZ to DLE scaling is expected to take place at a time of
order $t_c$ because the changes in the growth rate and in the roughness scaling
are caused by the same mechanisms. In other models with crossover between
different growth dynamics, it is observed that matching the scaling relations
for the roughness of both dynamics leads to correct predictions of $t_c$
\cite{chamereis,rdcor}. Following this reasoning, the $W$ values in Eqs.
(\ref{wkpz}) and (\ref{wdle}) must be the same at time $t_c\sim 1/p^2$ and for
a height $h_c\sim 1/p$. This gives
\begin{equation}
B\sim p^{-2/3} / \left( C_1-C_2\ln{p}\right) ,
\label{eqb}
\end{equation}
where $C_1$ and $C_2$ are constants.
In order to test relation (\ref{eqb}), we plot $p^{-2/3}/B$ as a function of
$\ln{p}$ in Fig. 7. We obtain a reasonable linear fit of the data for three
orders of magnitude of $p$, which suggests the validity of our approach.

The crossover of surface roughness is slightly different in $2+1$ dimensions
because the exponent $\beta$ of the KPZ class is changed \cite{kpz2d} and the
roughness of DLE saturates at a finite value \cite{krug}. However, the model
rules indicate that the short and long time dynamics are also KPZ and Laplacian
growth, respectively.

\section{Summary and Conclusion}

We studied the crossover from Kardar-Parisi-Zhang growth to Laplacian front
growth in a model for corrosion and passivation of a metalic surface. At small
times, a regime of constant
growth of the corrosion front is observed, with the roughness obeying KPZ
scaling. At long times, a large fraction of the layer is blocked,
leading to a global corrosion rate that decreases as $t^{-1/2}$. The roughness
shows logarithmic scaling, with amplitudes depending on the corrosion
probability $p$. Scaling arguments show that the crossover between these
different dynamics takes place at $t_c\sim 1/p^2$, and that the height of the
corrosion front at that time is $h_c\sim 1/p$. Universal scaling relations
between the height of the corrosion front $h$ and the time $t$ are obtained, as
well as relations for the surface roughness. All results are confirmed by
numerical simulations in square lattices, although the scaling arguments are
also valid in three-dimensions.

As far as we know, the crossover from KPZ to DLE scaling in statistical growth
models was not predicted before. The inverse situation, i. e. the crossover from
DLE (short times) to KPZ (long times) was suggested to appear in an extension
of the DLE model where the movement of the eroding particles is biased towards
the surface \cite{krug}. However, recent simulation results suggest a crossover
to Edwards-Wilkinson scaling instead of KPZ \cite{kim}. We also recall
that the crossover of our model must not be confused with the widely studied
crossover from KPZ to diffusion-limited aggregation (DLA) - see e. g. Ref.
\protect\cite{meakin,castro,zeguilherme2005,perez}. Indeed, surface
roughness scaling is very different in DLE and DLA.

The universal behavior discussed here
may be observed in real systems where the basic features illustrated in Fig. 1
are expected. Such tests are certainly possible with the
modern techniques of microscopy, that allow the determination of surface
roughness with good accuracy. One of the interesting points for comparison is
the time behavior of the roughness, which evolves from a rapid growth, typical
of KPZ scaling, to a very slow growth, typical of the logarithmic scaling in
Laplacian front growth. Moreover, it would also be interesting to confirm the
association of the changes in the growth rate with the changes in roughness
scaling.

\acknowledgements

FDAAR thanks the Institute of Physical Chemistry of the Polish
Academy of Sciences, where part of this work was done, for the
hospitality, and acknowledges support by CNPq (Brazil).


\vfill\eject

\begin{figure}[!h]
\includegraphics[clip,width=\textwidth, 
angle=0]{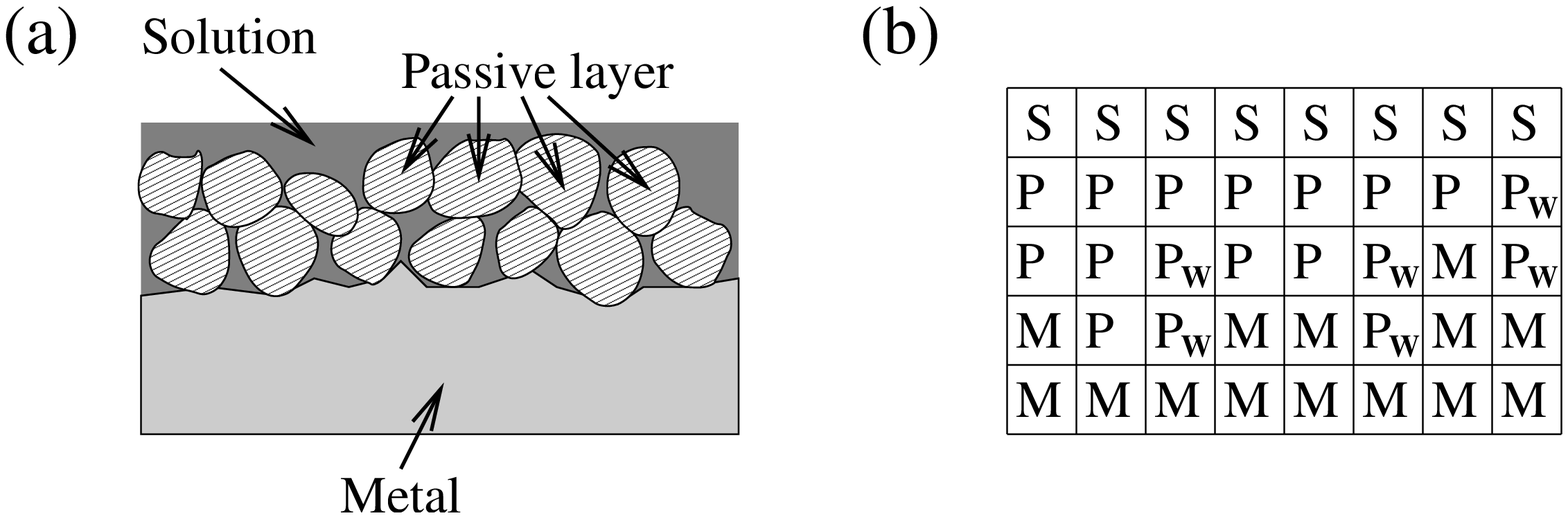}   
\caption{\label{fig1} (a) Scheme of the metal and the porous passive layer in
contact with the solution. (b) Lattice model after some steps of the corrosion
process.}
\end{figure}

\begin{figure}[!h]
\includegraphics[clip,width=0.9\textwidth, 
angle=0]{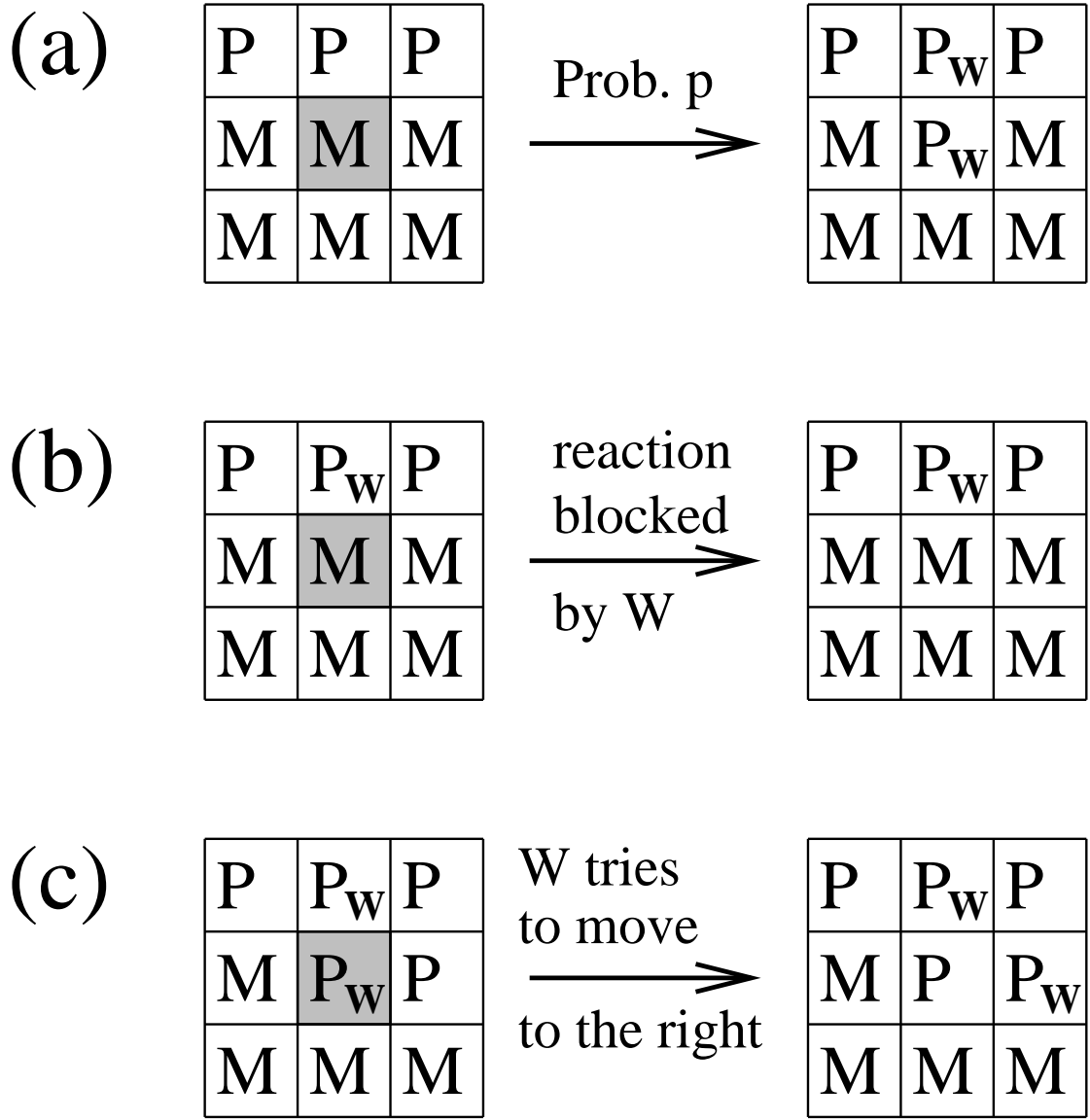}   
\caption{\label{fig2} (a) Corrosion of the central M site, which is passivated
(reaction $M\to P$) with the creation of two new W particles. (b) Example of an
M site (central site) whose passivation is not possible due to the blockage (by
W particle) of the only neighboring P (above the M site). (c) Example of
movement of a W particle, initially at the central P site.}
\end{figure}

\begin{figure}[!h]
\includegraphics[clip,width=\textwidth, 
angle=0]{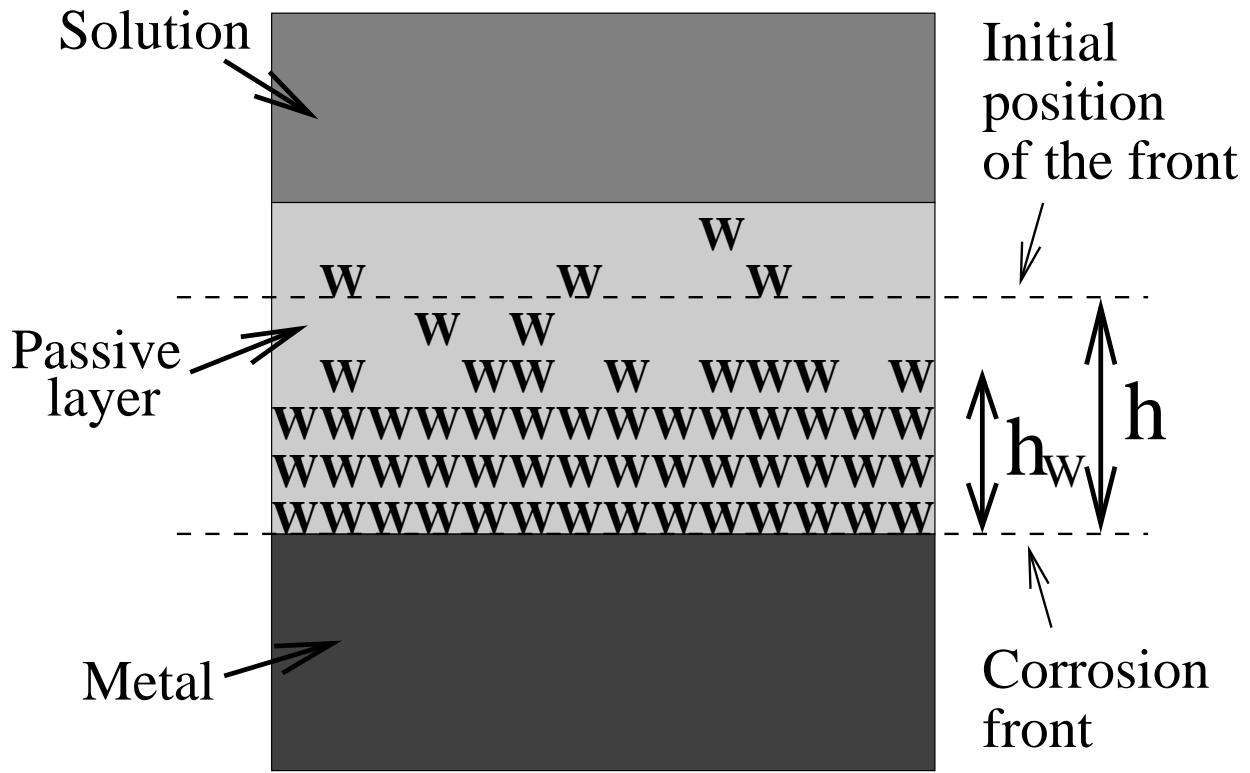}   
\caption{\label{fig3} Scheme of the lattice model after a long time, when a
large number of W particles are accumulated above the corrosion front, which
separates the metal and the passive layer. For simplicity, flat interfaces
between metal, passive layer and solution were drawn.}
\end{figure}

\begin{figure}[!h]
\includegraphics[clip,width=\textwidth, 
angle=0]{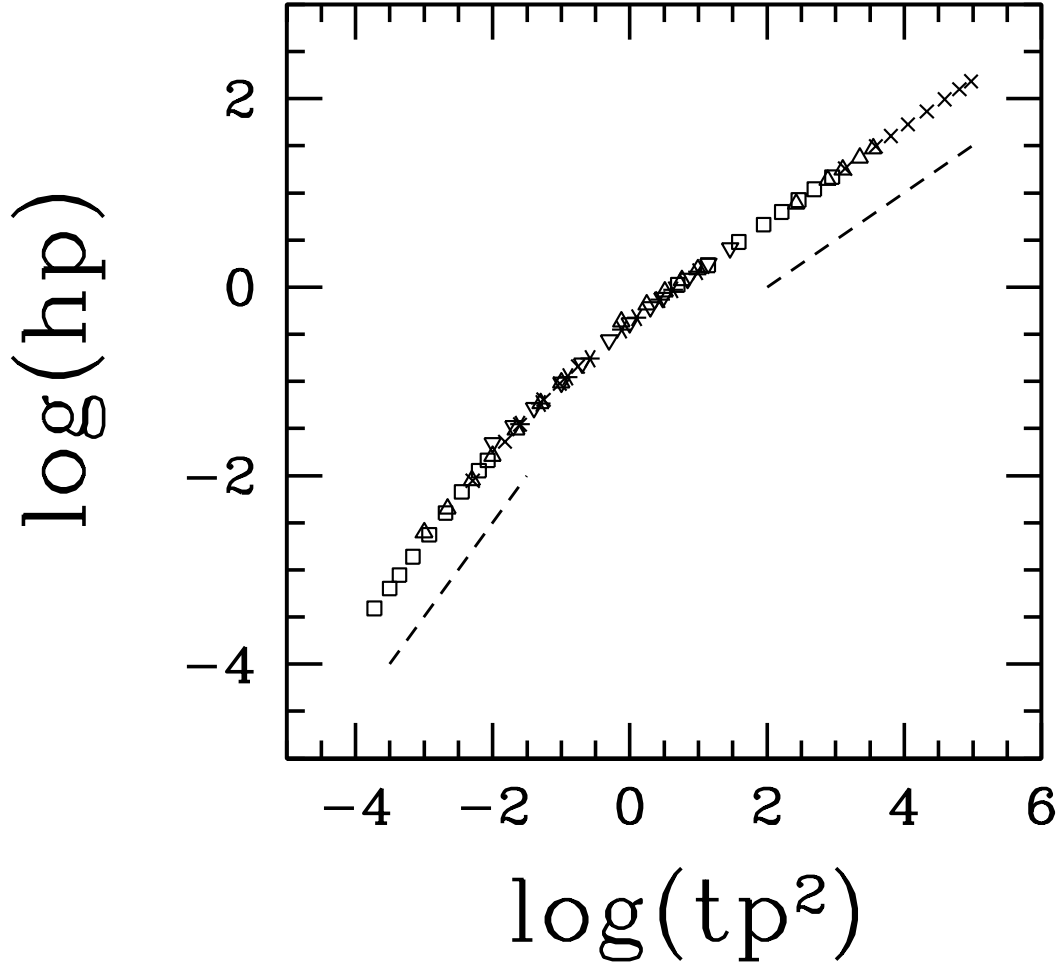}   
\caption{\label{fig4} Scaling plot of $h/h_c \sim hp$ versus $t/t_c\sim tp^2$
with data for $D=1$ and $p=0.5$ (crosses, right side), $p=0.1$ (up triangles,
right side), $p=0.05$ (squares, right side), $p=0.01$ (down triangles),
$p=0.005$ (asterisks), $p=0.001$ (up triangles, left side), $p=0.0005$
(crosses, left side), $p=0.00005$ (squares, left side). The dashed lines have
slopes $1$ (left) and $1/2$ (right).}
\end{figure}

\begin{figure}[!h]
\includegraphics[clip,width=\textwidth, 
angle=0]{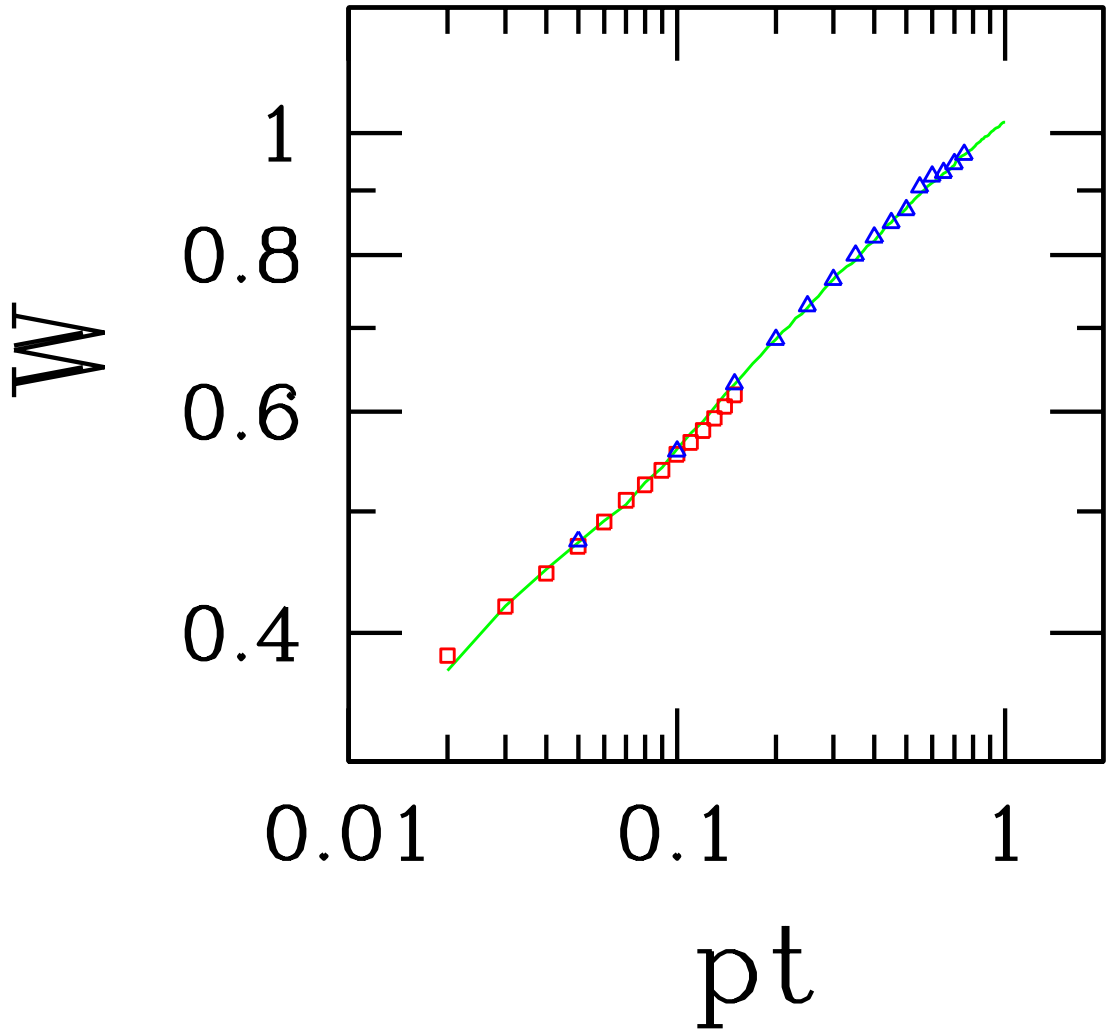}   
\caption{\label{fig5} Short-time evolution of the surface roughness, with time
scaling accounting only for the local corrosion rate $p$. Data for $p=0.001$
(solid line), $p=0.01$ (squares) and $p=0.00005$ (triangles) are shown.}
\end{figure}

\begin{figure}[!h]
\includegraphics[clip,width=\textwidth, 
angle=0]{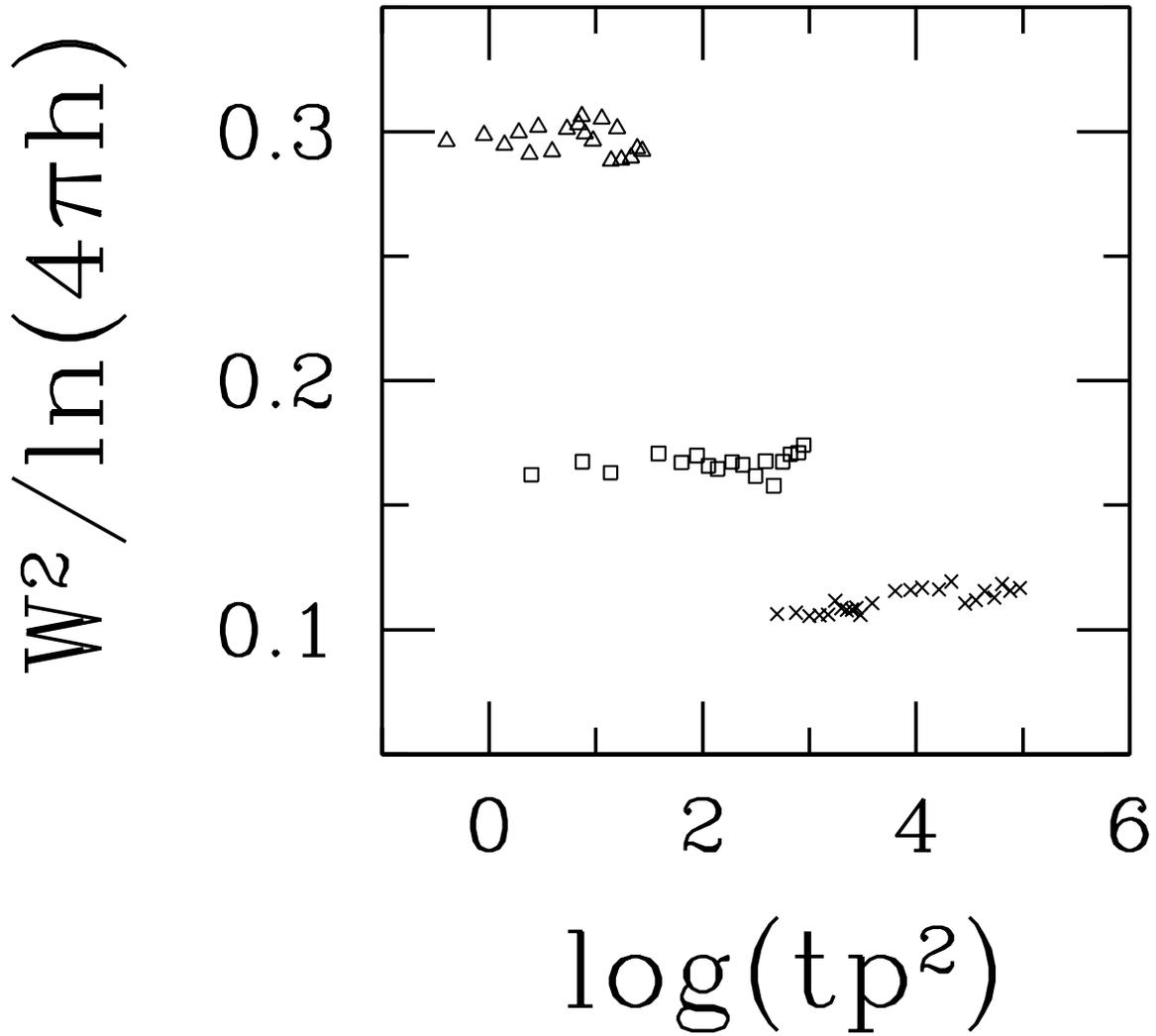}   
\caption{\label{fig6} Scaling plot of the roughness in the DLE regime (long
times): $p=0.5$ (crosses), $p=0.05$ (squares) and $p=0.01$ (triangles).}
\end{figure}

\begin{figure}[!h]
\includegraphics[clip,width=\textwidth, 
angle=0]{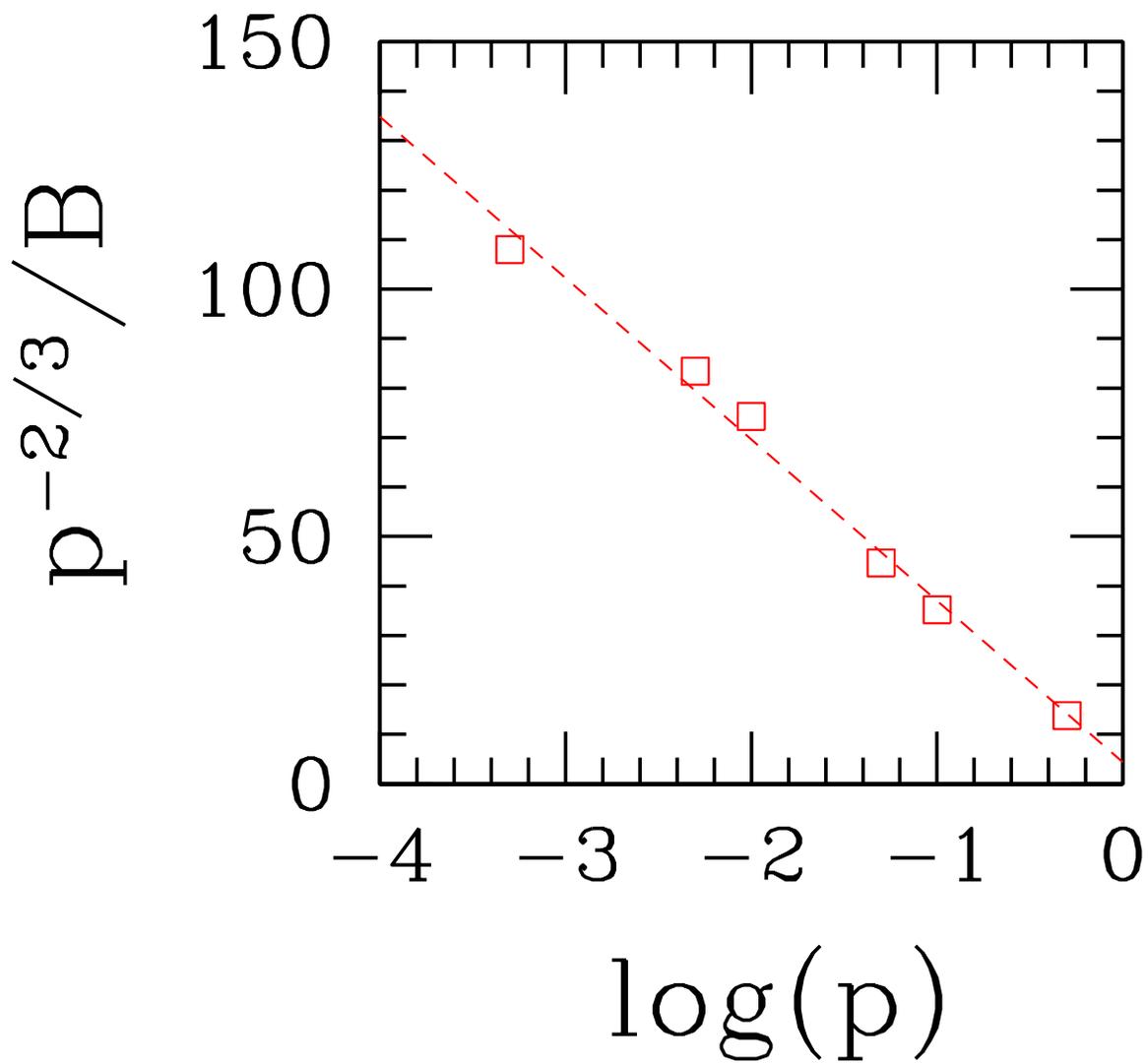}   
\caption{\label{fig7} Ratio $p^{-2/3}/B$ as a function of $\log{p}$, obtained
from long time data for values of $p$ ranging from $0.5$ to $0.0005$.}
\end{figure}

\end{document}